\documentclass[a4paper]{article}
\RequirePackage[english]{babel}
\RequirePackage[latin1]{inputenc}
\RequirePackage[T1]{fontenc}
\RequirePackage{mathrsfs}
\RequirePackage{amsmath}
\RequirePackage{amssymb}
\RequirePackage{amsbsy}
\RequirePackage{bm}
\begin{document}
\title{\bf{Hadronic Electroweak\\ Spin-Torsion Interactions}}
\author{Luca Fabbri\\ \footnotesize  Dipartimento di Fisica, Universit\`a di Bologna, 
Via Irnerio 46, 40126 Bologna, ITALY}
\date{}
\maketitle
\begin{abstract}
In a previous paper we considered the most general field equations for a system of two fermions of which one single-handed, showing that the spin-torsion interactions among these spinors had the same structure of the electroweak forces among leptons; in this paper we consider the most general field equations for a system of two fermions, showing that the spin-torsion interactions among these spinors have a structure similar to that of the electroweak forces among hadrons: possible differences are discussed, and consequently further extensions are speculated.
\end{abstract}
\section*{Introduction}
In the structure of the Dirac field equation given for the most general fermionic dynamics, the most general spinorial derivative contains torsion; since torsion is a tensor then all torsional contribution can be separated away: the most general spinorial derivative with torsion is thus decomposed in terms of the simplest spinorial derivative without torsion plus torsional contributions. Eventually when the field equations coupling torsion to the spin density are used, these torsional contributions get the form of specific spinorial autointeractions.

In the case in which many spinors are considered, then the spin density of the system is the sum of all spin densities of each spinor involved; thus the additional interactions display both the form of spinorial autointeraction of any spinor with itself and spinorial mutual interactions of each spinor with every other reciprocally. In the case of a system of two spinors of which one is a spinor and the other a semi-spinor left-handed these spinorial interactions have a structure that is identical to that of the leptonic weak forces \cite{f}.

Instead in the present paper we consider the system of two spinors to prove that the spinorial interactions have a structure that is similar to that of the hadronic weak interactions; therefore, unlike in the previous paper where the interactions were identical, here the interactions will only be similar. Eventually, we will discuss some consequences of these discrepancies.
\section{Torsional interaction}
As in the previous paper we consider a set of $k$ matter fields labelled with the indices in Latin each of which governed by the matter field equation
\begin{eqnarray}
&i\gamma^{\mu}D_{\mu}\phi^{a}=0
\label{matter}
\end{eqnarray}
given in the massless case, and these equations come along with the background field equations given for the combination of the Ricci and Cartan tensors in terms of the energy and the spin distribution of the matter field as
\begin{eqnarray}
&G_{\alpha\beta}=\frac{i}{4}\sum_{a}\left[\bar{\psi}^{a}\gamma_{\alpha}D_{\beta}\psi^{a}
-D_{\beta}\bar{\psi}^{a}\gamma_{\alpha}\psi^{a}\right]
\label{metric}
\end{eqnarray}
and
\begin{eqnarray}
&Q_{\mu\alpha\beta}
=-\frac{i}{4}\sum_{a}\bar{\psi}^{a}\{\gamma_{\mu},\sigma_{\alpha\beta}\}\psi^{a}
\label{torsion}
\end{eqnarray}
according to the prescription of the Einstein-Sciama--Kibble scheme; then it is possible to use the torsion as given by the field equations (\ref{torsion}) in order to substitute torsion with the spin of the spinor fields as
\begin{eqnarray}
&i\gamma^{\mu}\nabla_{\mu}\psi^{a}
+\frac{3}{16}\sum_{b}\bar{\psi}^{b}\gamma_{\mu}\gamma\psi^{b}
\gamma^{\mu}\gamma\psi^{a}=0,\ \ \ \ a=1\hdots k
\label{matterfield}
\end{eqnarray}
where we see that spinorial bilinears appear, exactly as they did in reference \cite{f}.

\subsection{Torsional interaction:\\ spin coupling of two spinors}
In this paper we want to compare these spinorial interactions with the hadronic weak forces: thus it is necessary that the two systems have the same spinorial field content, and consequently we are going to consider the case in which two spinor fields are present, and where they have both projections.

So in the case we have two spinors the matter field equations are given by the matter field equations (\ref{matterfield}) with $k=2$ and they can be explicitly written as
\begin{eqnarray}
&i\gamma^{\mu}\nabla_{\mu}\psi^{1}
+\frac{3}{16}\bar{\psi}^{1}\gamma_{\mu}\gamma\psi^{1}\gamma^{\mu}\gamma\psi^{1}
+\frac{3}{16}\bar{\psi}^{2}\gamma_{\mu}\gamma\psi^{2}\gamma^{\mu}\gamma\psi^{1}=0\\
&i\gamma^{\mu}\nabla_{\mu}\psi^{2}
+\frac{3}{16}\bar{\psi}^{1}\gamma_{\mu}\gamma\psi^{1}\gamma^{\mu}\gamma\psi^{2}
+\frac{3}{16}\bar{\psi}^{2}\gamma_{\mu}\gamma\psi^{2}\gamma^{\mu}\gamma\psi^{2}=0
\label{matterfields}
\end{eqnarray}
as it can be seen by separating the fields.

Moreover, we can separate the right-handed and left-handed projections as
\begin{eqnarray}
\nonumber
&i\gamma^{\mu}\nabla_{\mu}\psi^{1}_{L}
+\frac{3}{16}\bar{\psi}^{1}_{L}\gamma_{\mu}\psi^{1}_{L}\gamma^{\mu}\psi^{1}_{L}
+\frac{3}{16}\bar{\psi}^{2}_{L}\gamma_{\mu}\psi^{2}_{L}\gamma^{\mu}\psi^{1}_{L}-\\
&-\frac{3}{16}\bar{\psi}^{1}_{R}\gamma_{\mu}\psi^{1}_{R}\gamma^{\mu}\psi^{1}_{L}
-\frac{3}{16}\bar{\psi}^{2}_{R}\gamma_{\mu}\psi^{2}_{R}\gamma^{\mu}\psi^{1}_{L}=0\\
\nonumber
&i\gamma^{\mu}\nabla_{\mu}\psi^{1}_{R}
-\frac{3}{16}\bar{\psi}^{1}_{L}\gamma_{\mu}\psi^{1}_{L}\gamma^{\mu}\psi^{1}_{R}
-\frac{3}{16}\bar{\psi}^{2}_{L}\gamma_{\mu}\psi^{2}_{L}\gamma^{\mu}\psi^{1}_{R}+\\
&+\frac{3}{16}\bar{\psi}^{1}_{R}\gamma_{\mu}\psi^{1}_{R}\gamma^{\mu}\psi^{1}_{R}
+\frac{3}{16}\bar{\psi}^{2}_{R}\gamma_{\mu}\psi^{2}_{R}\gamma^{\mu}\psi^{1}_{R}=0\\
\nonumber
&i\gamma^{\mu}\nabla_{\mu}\psi^{2}_{L}
+\frac{3}{16}\bar{\psi}^{1}_{L}\gamma_{\mu}\psi^{1}_{L}\gamma^{\mu}\psi^{2}_{L}
+\frac{3}{16}\bar{\psi}^{2}_{L}\gamma_{\mu}\psi^{2}_{L}\gamma^{\mu}\psi^{2}_{L}-\\
&-\frac{3}{16}\bar{\psi}^{1}_{R}\gamma_{\mu}\psi^{1}_{R}\gamma^{\mu}\psi^{2}_{L}
-\frac{3}{16}\bar{\psi}^{2}_{R}\gamma_{\mu}\psi^{2}_{R}\gamma^{\mu}\psi^{2}_{L}=0\\
\nonumber
&i\gamma^{\mu}\nabla_{\mu}\psi^{2}_{R}
-\frac{3}{16}\bar{\psi}^{1}_{L}\gamma_{\mu}\psi^{1}_{L}\gamma^{\mu}\psi^{2}_{R}
-\frac{3}{16}\bar{\psi}^{2}_{L}\gamma_{\mu}\psi^{2}_{L}\gamma^{\mu}\psi^{2}_{R}+\\
&+\frac{3}{16}\bar{\psi}^{1}_{R}\gamma_{\mu}\psi^{1}_{R}\gamma^{\mu}\psi^{2}_{R}
+\frac{3}{16}\bar{\psi}^{2}_{R}\gamma_{\mu}\psi^{2}_{R}\gamma^{\mu}\psi^{2}_{R}=0
\label{fundamentalmatterfields}
\end{eqnarray}
in which all spinors are expressed in their left-handed or the right-handed decompositions written in their irreducible chiral representation.

By using the Fierz identities we can write these field equations in formally equivalent ways that will be more suited for the task we want to pursue, but whereas in \cite{f} this rearrangements was somehow quite natural, here the rearrangement has to take into account an additional factor. In the case of leptons, we had that the field content was given by two left-handed fields and a single right-handed field, so that in principle we could have postulated that no abelian charge were to be present, and still we would have had the doublet of left-handed fields mixing leaving the singlet of right-handed field alone, as expected; in the case of hadrons, we have that the field content is given by two left-handed and two right-handed fields, so that if in principle we postulated that no abelian charge were to be present, then we would have both the doublet of left-handed and right-handed fields mixing, unlike what is expected: to avoid this, we have to let the abelian charges be present, postulating that the two left-handed fields have equal abelian charges and the two right-handed fields have different abelian charges, so that the doublet of left-handed fields would still mix leaving the two singlets of right-handed fields alone, as expected. However, if we let abelian charges then also an abelian field must be introduced, which will be a fundamental field with independent degrees of freedom on its own.

\paragraph{Massless fundamental hadrons and composite scalar and vector fields.} So far we have started from a system of field equations for two spinors writing them in the equivalent form above, which may itself be written now as
\begin{eqnarray}
&i\gamma^{\mu}\nabla_{\mu}L
-\frac{1}{2}g\vec{\sigma}\cdot\vec{A}_{\mu}\gamma^{\mu}L
-\frac{1}{6}g'B_{\mu}\gamma^{\mu}L-G_{u}i\sigma^{2}\phi^{\ast}u-G_{d}\phi d\approx0\\
&i\gamma^{\mu}\nabla_{\mu}u-\frac{2}{3}g'B_{\mu}\gamma^{\mu}u+G_{u}i\phi^{T}\sigma^{2}L\approx0\\
&i\gamma^{\mu}\nabla_{\mu}d+\frac{1}{3}g'B_{\mu}\gamma^{\mu}d-G_{d}\phi^{\dagger}L\approx0
\label{equivalentfundamentalmatterfields}
\end{eqnarray}
in which the expected supplementary interactions were three-field vertices therefore negligible, and the system of field equations thus approximated has a form known well. Indeed, in the leading order, this is the form of the field equations for the quark fields before the symmetry breaking in the standard model.

In order to be able to get this form we rename the spinor fields
\begin{eqnarray}
&\left(\psi^{1}_{R}\right)=u\ \ \ \ \left(\psi^{2}_{R}\right)=d\ \ \ \ \ \ \ \ 
\left(\begin{tabular}{c}$\psi^{1}_{L}$\\ $\psi^{2}_{L}$\end{tabular}\right)=L
\end{eqnarray}
as new quark fields: then we have to consider their bilinear fields defining
\begin{eqnarray}
&\frac{15}{32G_{d}}\left(\bar{d}L\right)
+\frac{3i}{16G_{u}}\left(\bar{u}\sigma^{2}L\right)^{\ast}=\phi
\end{eqnarray}
for the scalar field; and finally we have
\begin{eqnarray}
&\frac{9}{32}\left(\bar{L}\gamma_{\mu}\frac{\mathbb{I}}{2}L
+2\bar{u}\gamma_{\mu}u-\bar{d}\gamma_{\mu}d\right)+q\Gamma_{\mu}=g'B_{\mu}\\
&\frac{9}{32}\bar{L}\gamma_{\mu}\frac{\vec{\sigma}}{2}L=g\vec{A}_{\mu}
\end{eqnarray}
for the vector fields. As discussed above, the abelian field $\Gamma_{\mu}$ must be fundamental and its coupling is given in terms of the $-\frac{1}{6}q$, $-\frac{2}{3}q$ and $\frac{1}{3}q$ charges for the left-handed $L$, right-handed $u$ and $d$ quarks respectively; the abelian field given by the current $\frac{9}{32}(\bar{L}\gamma_{\mu}\frac{\mathbb{I}}{2}L +2\bar{u}\gamma_{\mu}u -\bar{d}\gamma_{\mu}d)$ is composite in terms of those quarks and what allows the present prescription to work is that this current is defined once but its appearance occurs in terms of the $-\frac{1}{6}$, $-\frac{2}{3}$ and $\frac{1}{3}$ factors in front of the $L$, $u$ and $d$ quarks respectively: this is the reason for which the current $\frac{9}{32} (\bar{L}\gamma_{\mu}\frac{\mathbb{I}}{2}L+2\bar{u}\gamma_{\mu}u-\bar{d}\gamma_{\mu}d)$ and the abelian field $q\Gamma_{\mu}$ may always be reabsorbed into the definition of the abelian field $g'B_{\mu}$ still fundamental and still occurring with the $-\frac{1}{6}$, $-\frac{2}{3}$ and $\frac{1}{3}$ factors in front of the $L$, $u$ and $d$ quarks again respectively. By following the same method presented in \cite{f}, it is possible to establish the form of the transformation laws for the quarks, the scalar and vector fields before the symmetry breaking in the standard model.

That torsional interactions and weak forces may be linked was discussed for the leptons \cite{f,s-s,h-h-k-n,s-s/1,h-d,s-g,s-s/2}, but never for the quarks.

\subparagraph{Massless fundamental hadrons and composite scalar and vector fields: structure of $U(1)\times SU(2)_{L}$ local electroweak gauge interaction.} Finally we consider the above field equations written as
\begin{eqnarray}
&i\gamma^{\mu}\mathbb{D}_{\mu}L-G_{u}i\sigma^{2}\phi^{\ast}u-G_{d}\phi d\approx0\\
&i\gamma^{\mu}\mathbb{D}_{\mu}u+G_{u}i\phi^{T}\sigma^{2}L\approx0\\
&i\gamma^{\mu}\mathbb{D}_{\mu}d-G_{d}\phi^{\dagger}L\approx0
\label{invariantequivalentfundamentalmatterfields}
\end{eqnarray}
in which the derivatives have been written in a compact form.

This compact form is obtained upon definition of the derivatives
\begin{eqnarray}
&\mathbb{D}_{\mu}L=\nabla_{\mu}L
+\frac{i}{2}(g\vec{\sigma}\cdot\vec{A}_{\mu}+\frac{1}{3}g'B_{\mu})L\\
&\mathbb{D}_{\mu}u=\nabla_{\mu}u+\frac{2i}{3}g'B_{\mu}u\\
&\mathbb{D}_{\mu}d=\nabla_{\mu}d-\frac{i}{3}g'B_{\mu}d
\end{eqnarray}
covariant for general $U(1)\times SU(2)_{L}$ local transformations. Again it is important to stress that the abelian part is constituted by both fundamental and composite fields whereas the non-abelian part is entirely built in terms of composite fields, and the generalization to local transformation is possible since the gauge field is local by definition and the massless $d$ and $u$ and also $L$ quarks are functions of the spacetime so that their mixing may take place with coefficients depending on the spacetime themselves. In this way we have defined the covariant derivatives of quark fields before the symmetry breaking in the standard model.

Now by following the same procedure we have followed in \cite{f}, it is possible to see that the mass of the scalar field will depend on the mass of the quarks we are considering; differently from the case of leptons however, here there is a quark possessing a mass that is high enough to give to the scalar field a mass larger than the Linde-Weinberg lower bound \cite{l,w}, with the consequence that the contribution of the effective potential determining the spontaneous breakdown of the symmetry are larger than the contributions that add up when all radiative corrections are accounted: in this way the stability of the vacuum is ensured.

Finally we remark that after the Higgs potential has induces the symmetry breaking and that from the Higgs field three degrees of freedom have been transferred to three massless vector fields to make them massive vector fields, we are left with one massless vector field, being a combination of composite quark currents and fundamental degrees of freedom; as we move from the Fermi scale to larger scales, the composite quark current would vanish leaving only the fundamental field, with the consequence that in the limit of lower energies, only the independent degrees of freedom would be preserved, compatibly with the results found in \cite{b}, where it is proved that so long as a perturbative approach is pursued, the photon appears as an elementary field. This abelian field is then able to propagate over longer distances, compatibly with what the electrodynamic field is expected to do. In this situation, for large scales the electrodynamics we observe is recovered, but at the Fermi scale electrodynamics behaves differently, both here due to the presence of the quark currents and in the standard model due to the presence of the weak bosons; on the other hand, here the quark currents build the weak bosons and in the standard model we have fundamental weak bosons, and this is what makes the substantial differences between the two approaches. As a consequence, except for the fundamental part of electrodynamics governed as usual by Maxwell equations, for weak interactions there are no dynamical field equations, precisely because these fields are composite, unlike in the standard model, where they are fundamental, and this fact might be relevant for the detection of possible discrepancies at the Fermi scale.

On the other hand however, at Fermi scales and beyond, the results we have obtained will in general depend also on the terms like the three-particle interactions that we have neglected, in which case we should expect even further discrepancies, and the hope is of course that such discrepancies will display themselves in the behaviour of quarks; the two problems we may face are, in a theoretical perspective, that although in principle it is possible to start from the known dynamics of quarks to deduce the dynamics of the composite weak bosons, in practice this will turn out to be quite difficult, and in the phenomenological perspective, the study of a complex system like the hadron itself, let alone the nucleus as a whole, will turn out to be even more difficult. Maybe in the future methods may be develop to help achieve this task.
\section*{Conclusion}
In this paper we have proved that the field equations for two spinors coupled through torsion and free of any other interaction are formally similar to the field equations for massless hadrons without torsion but with electroweak gauge interactions; contrary to the case of leptons, where there was a formal equivalence, for the case of hadrons, there is a formal equivalence only up to supplementary interactions appearing as three-field interactions: therefore although these extra terms may be negligible there are nevertheless differences. In the case in which these thee-field interactions are neglected, the resulting system of field equations has been used to discuss possible consequences for the stability properties of a possible composite Higgs boson and for the structure of possible composite weak bosons; but in general for the dynamics of nucleons there are discrepancies between the approach followed in this paper and that of the standard model, although they may be difficult to calculate explicitly and even more difficult to detect in experiments. At any rate, one of the problems that may affect this model is that of the energy scale, but compared to the case of leptons the case of hadrons is not changed and the same consideration drawn in \cite{f} are valid here.

A final point we wish to elucidate is that all we have discussed is based on taking into account Dirac fermions, employing torsion to reproduce the electroweak interactions before the symmetry breaking, while the situation may be very different if one considers ELKO fields \cite{a-g/1,a-g/2,a-l-s/1,a-l-s/2}, using torsion to give electroweak interactions prior the symmetry breaking phenomenon; in fact differently from the case of Dirac fields, for which torsion is algebraically related to the spin density, with ELKO fields, torsion is dynamically related to the spin density \cite{b/1,b/11}, and the procedure followed here to reproduce the electroweak interaction will be altered drastically. Although Dirac and ELKO fields may share structural similarities \cite{dr-hs,hs-dr}, nevertheless their dynamical differences are too profound to let electroweak interaction arise as effective fields.

\end{document}